\documentclass{kapproc} 

\usepackage{procps}

\usepackage[dvips]{graphicx}

\upperandlowercase

\kluwerbib 
\begin{document}

\articletitle[Structure Formation in dynamical Dark Energy models]
{Structure Formation in\\
dynamical Dark Energy models}

\author{A.V. Maccio'$^{1}$, S.A. Bonometto$^{1}$, R, Mainini$^{1}$
\& A. Klypin$^{2}$}

\affil{(1) Physics Dep. G. Occhialini, Univ. of Milano--Bicocca \& I.N.F.N., Sezione di Milano\\
Piazza della Scienza 2, 20126 Milano, Italy\\
(2) New Mexico State University, Las Cruces, New Mexico, USA
}
\email{maccio@mib.infn.it, bonometto@mib.infn.it, mainini@mib.infn.it,
aklypin@nmsu.edu}

\anxx{Dynamical Dark Energy}

\begin{abstract}
We perform n--body simulations for models with a DE component. 
Besides of DE with constant negative $w = p/\rho 
\geq -1$, we consider DE due to scalar fields, self--interacting
through RP or SUGRA potentials. According to our post--linear analysis,
at $z=0$, DM power spectra and halo mass functions do not depend on DE 
nature. This is welcome, as $\Lambda$CDM fits observations. 
Halo profiles, instead, are denser than $\Lambda$CDM.
For example, the density at 10$\, h^{-1}$kpc of a DE 
$\sim 10^{13}M_\odot$ halo exceeds $\Lambda$CDM by $\sim 40\, \%$. 
Differences, therefore, are small but, however, DE does not ease
the problem with cuspy DM profiles. On the contrary
it could ease the discrepancy between $\Lambda$CDM and
strong lensing data (Bertelmann 1998, 2002). 
We study also subhalos and find 
that, at $z=0$, the number of satellites coincides in all DE models.
At higher $z$, DE models show increasing differences from 
$\Lambda$CDM and among themselves; this is the obvious pattern to 
distinguish between different DE state equations.

\end{abstract}


\section{Introduction}
Deep survey and CBR data confirm that $\sim 70\, \%$ of the world is 
Dark Energy (see, e.g., {Efstathiou et al 2002, 
Percival et al 2002, Spergel et al 2003, Tegmark et al 2001, Netterfield
et al 2002, Pogosian et al 2003, Kogut et al 2003}), as needed to
have the accelerated expansion shown by SNIa data ({Riess et al 1998, 
Perlmutter et al 1999}). The nature of Dark Energy (DE) is a puzzle. 
$\Lambda$CDM needs a severe fine--tuning of vacuum energy.
DE with constant negative $w = p/\rho >-1 $ has
even less physical motivation. Apparently, the only viable 
alternative is dynamical 
DE, a {\it classical} self--interacting scalar field $\phi$ 
(Wetterich 1985). 
Among potentials $V(\phi)$ with a tracker solution, limiting
the impact of initial conditions,
Ratra--Peebles (1988, RP hereafter) and SUGRA (Brax \& Martin
1999, 2000) potentials bear a particle physics motivation.

Studying a dynamical DE model requires:
(i) a linear treatment, to yield
CBR spectra and transfer function; (ii) a post--linear treatment,
to yield halo virial density contrasts and mass functions; 
(iii) a non--linear treatment. 
Here we report results on (ii) and (iii).
We use the n--body program ART, modified
to deal with any dependence of $\Omega_m$ (matter density parameter)
on $a$ (scale factor). Mainini et al (2003b) give analytical fitting
formulae for such dependence. Further details 
are in Mainini, Maccio' \& Bonometto (2003a) and Klypin et al (2003).
RP and SUGRA are parametrized by the energy scale $\Lambda/{\rm GeV}$.

\section{Non--linear results}
Fig.~1 shows the evolution of the spectrum, as
obtained from simulations of $\Lambda$CDM and RP models 
($\Lambda$/GeV$=10^3$), the most distant models treated.
Models were normalized so to obtain the same number of halos
at $z=0$.

\begin{figure}[t]
\sidebyside
{\centerline{\includegraphics[width=2in,height=1.6in]{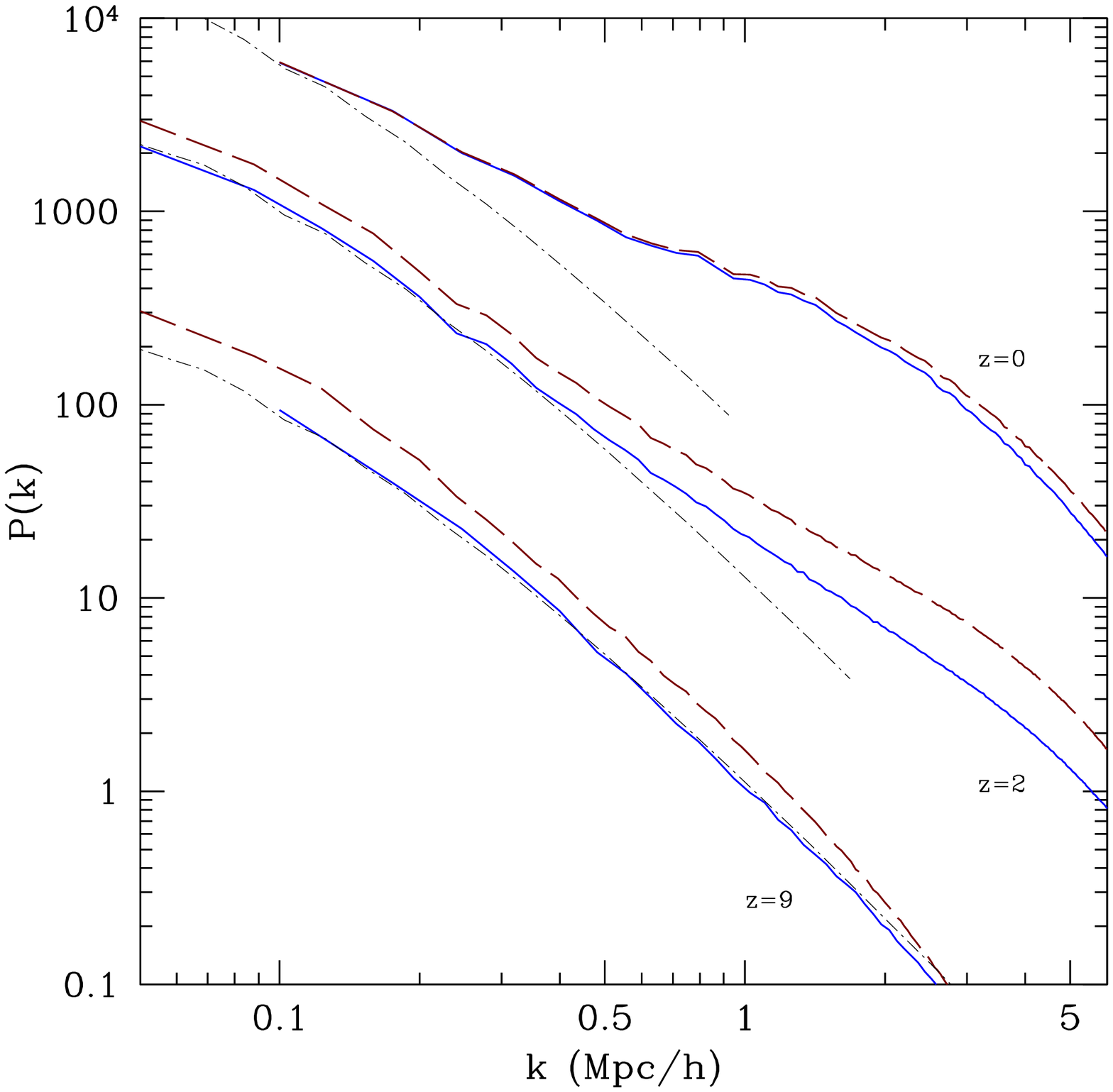}}
\caption{Spectrum evolution for $\Lambda$CDM (solid line) and RP (long
dashed); the dot--dashed line is the linear prediction for $\Lambda$CDM.}}
{\centerline{\includegraphics[width=2in,height=1.7in]{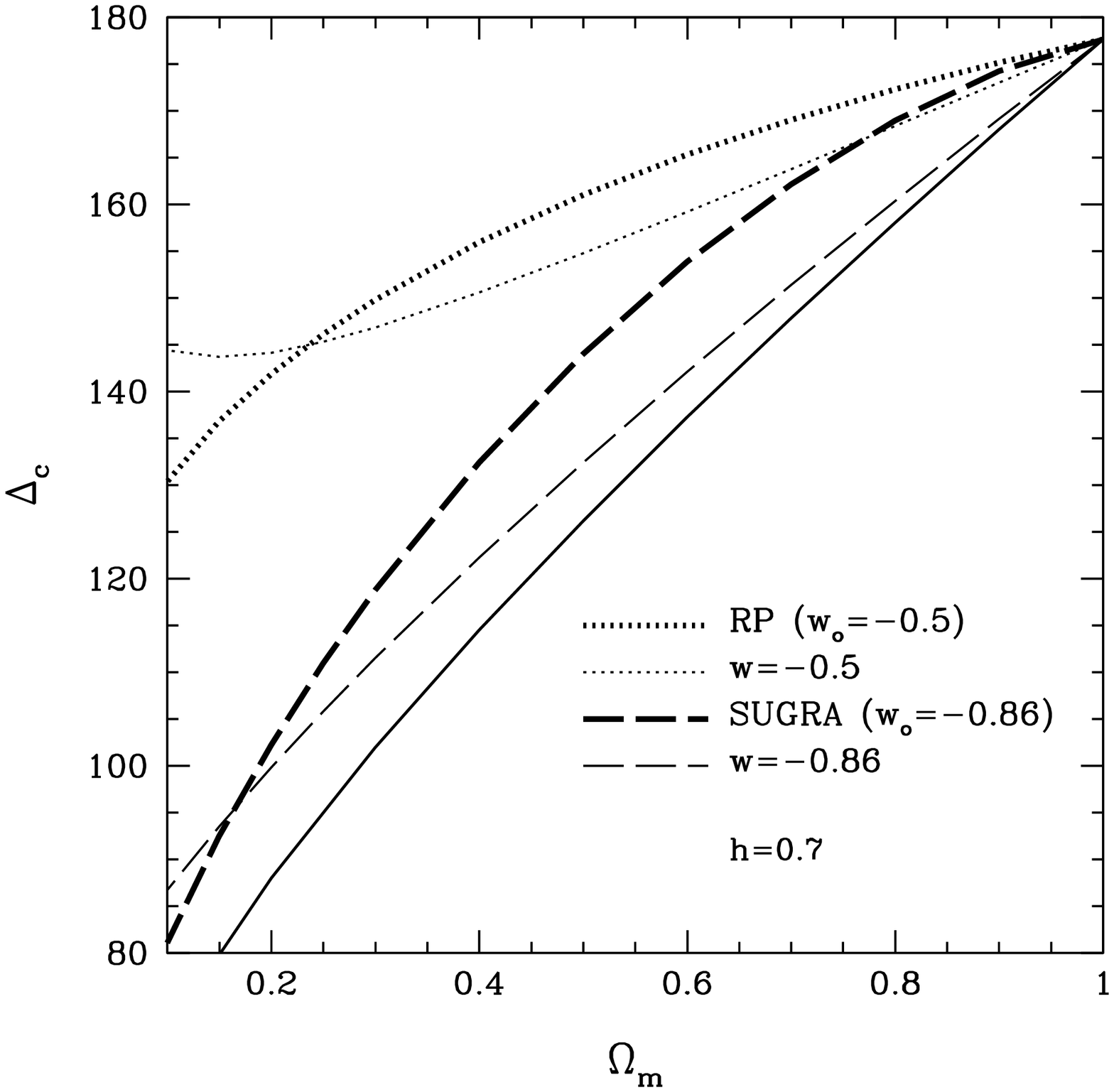}}
\caption{ The virial density contrast $\Delta_c$ vs. $\Omega_m$ at $z=0$.
Models are indicated in the frame }}
\end{figure}

Halos were extracted from simulations using the virial density contrasts
$\Delta_c$ obtained by Mainini et al (2003b), where one can find
plots for the dependence $\Delta_c(a)$; here we show $\Delta_c$ dependence
on $\Omega_m$ at $z=0$ (Fig.~2)
Figs. 3 \& 4 show the mass function and its evolution in a number of models.

\begin{figure}[t]
\sidebyside
{\centerline{\includegraphics[width=2in,height=1.8in]{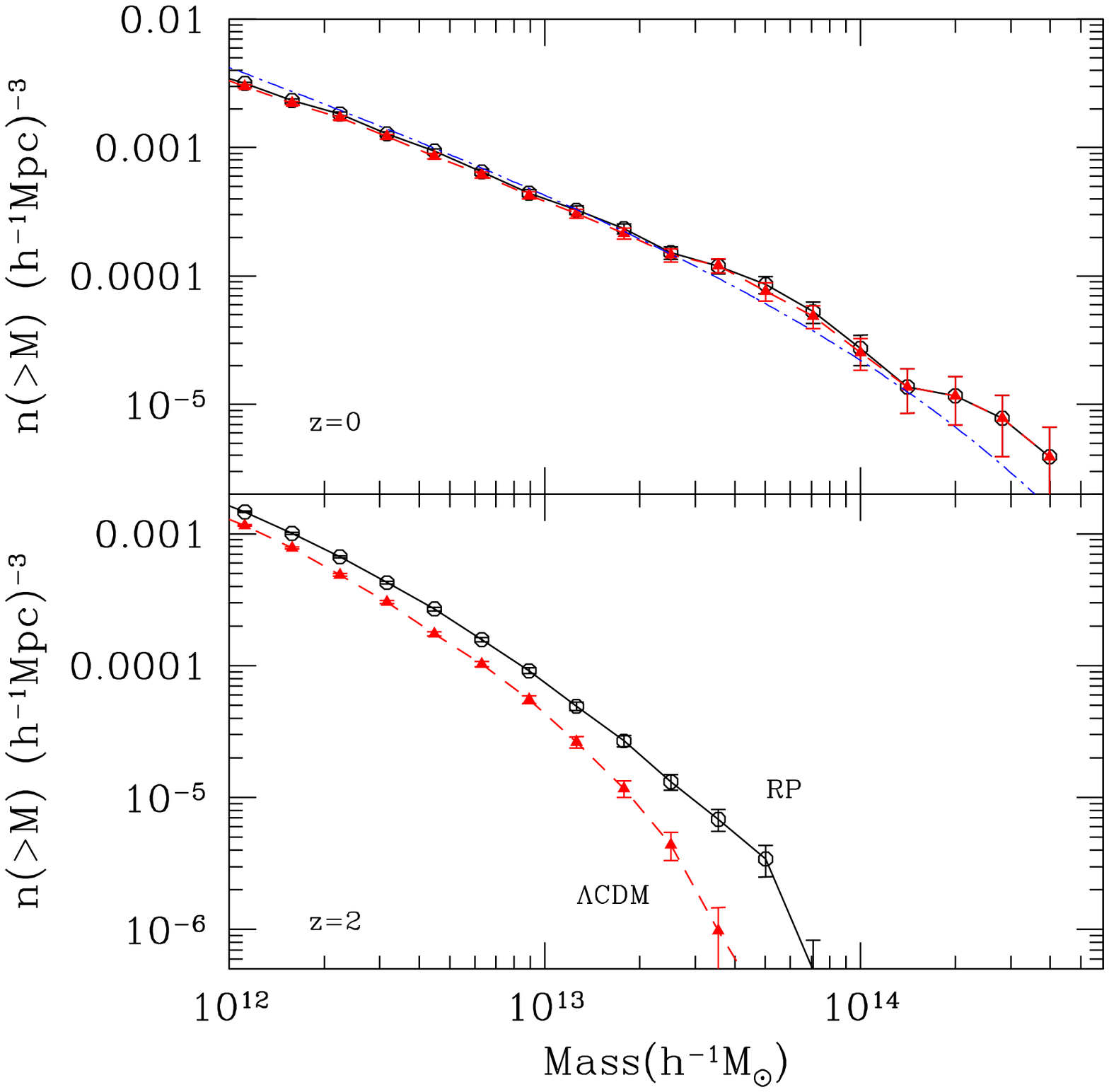}}
\caption{ Mass function at $z=0$ and $z=2$ for the same models of Fig.~1.
Evolution is faster for $\Lambda$CDM than for RP }}
{\centerline{\includegraphics[width=2in,height=1.8in]{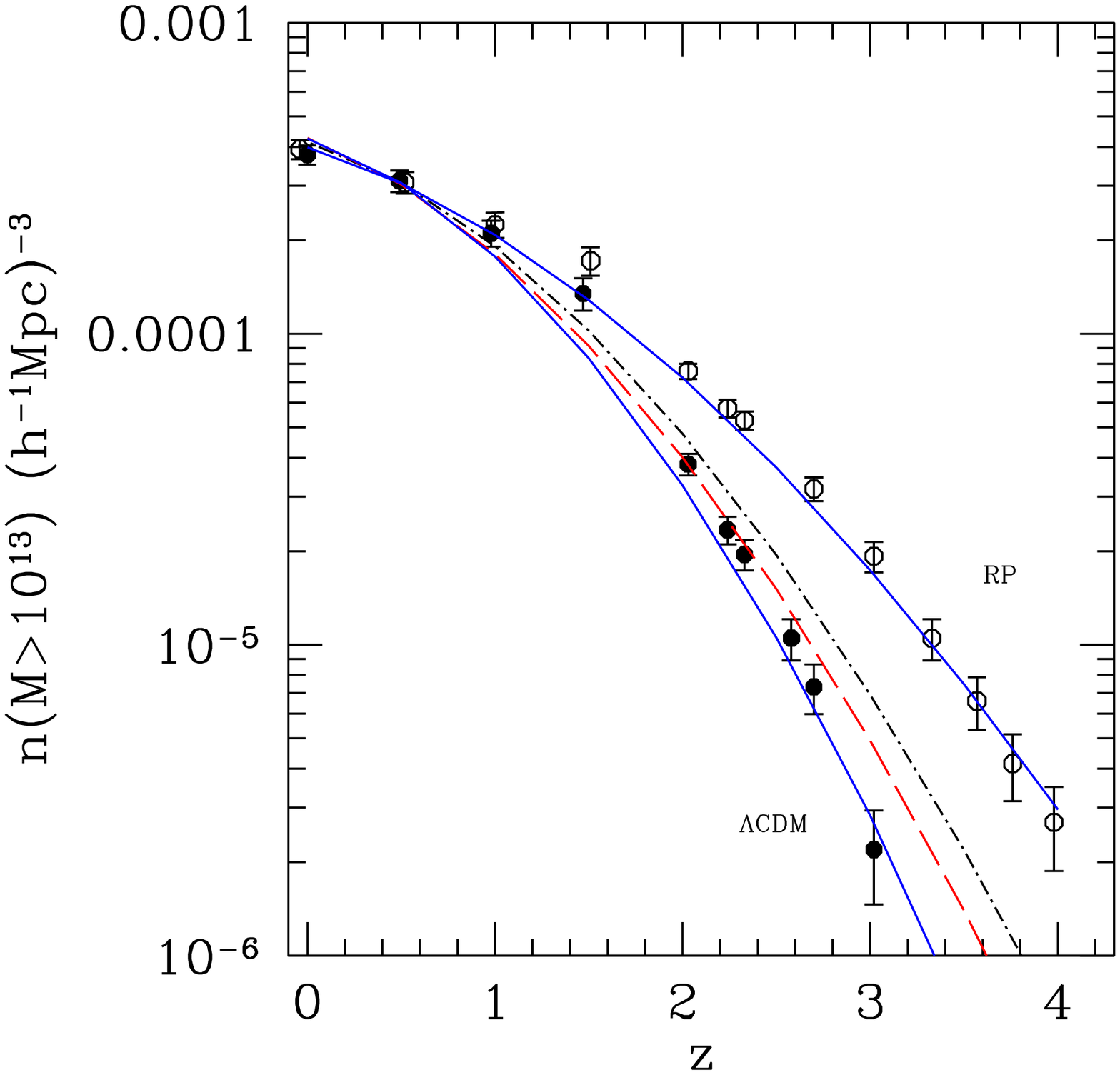}}
\caption{ Halo number evolution in various models.
Unlabeled curves refer to SUGRA and constant $w=-0.8$
 }}
\end{figure}

\begin{figure}[t]
\sidebyside
{\centerline{\includegraphics[width=2in,height=1.8in]{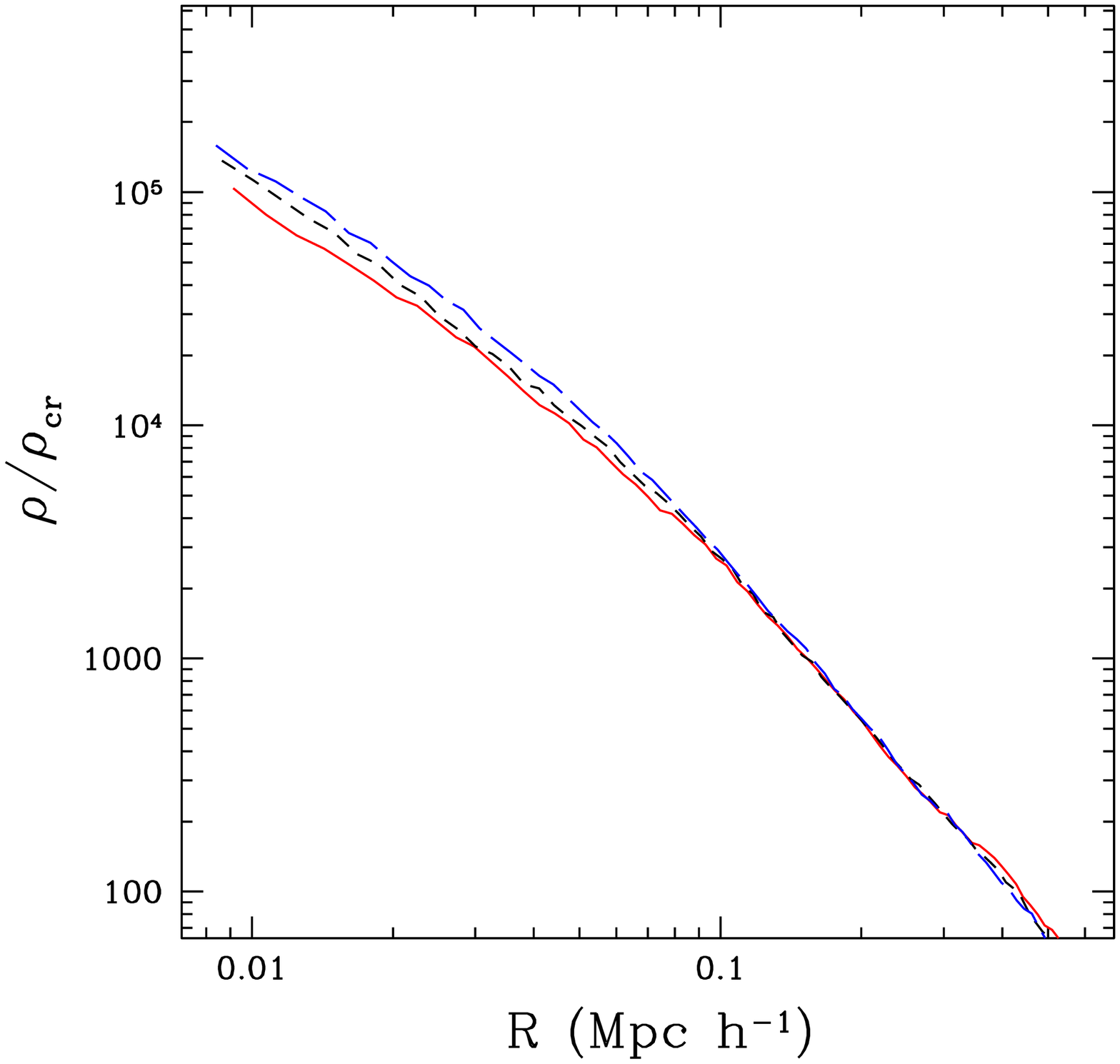}}
\caption{ Density profile for a single magnified halo.
Solid, short dashed, long dashed lines refer to $\Lambda$CDM,
SUGRA, RP. }}
{\centerline{\includegraphics[width=2in,height=1.8in]{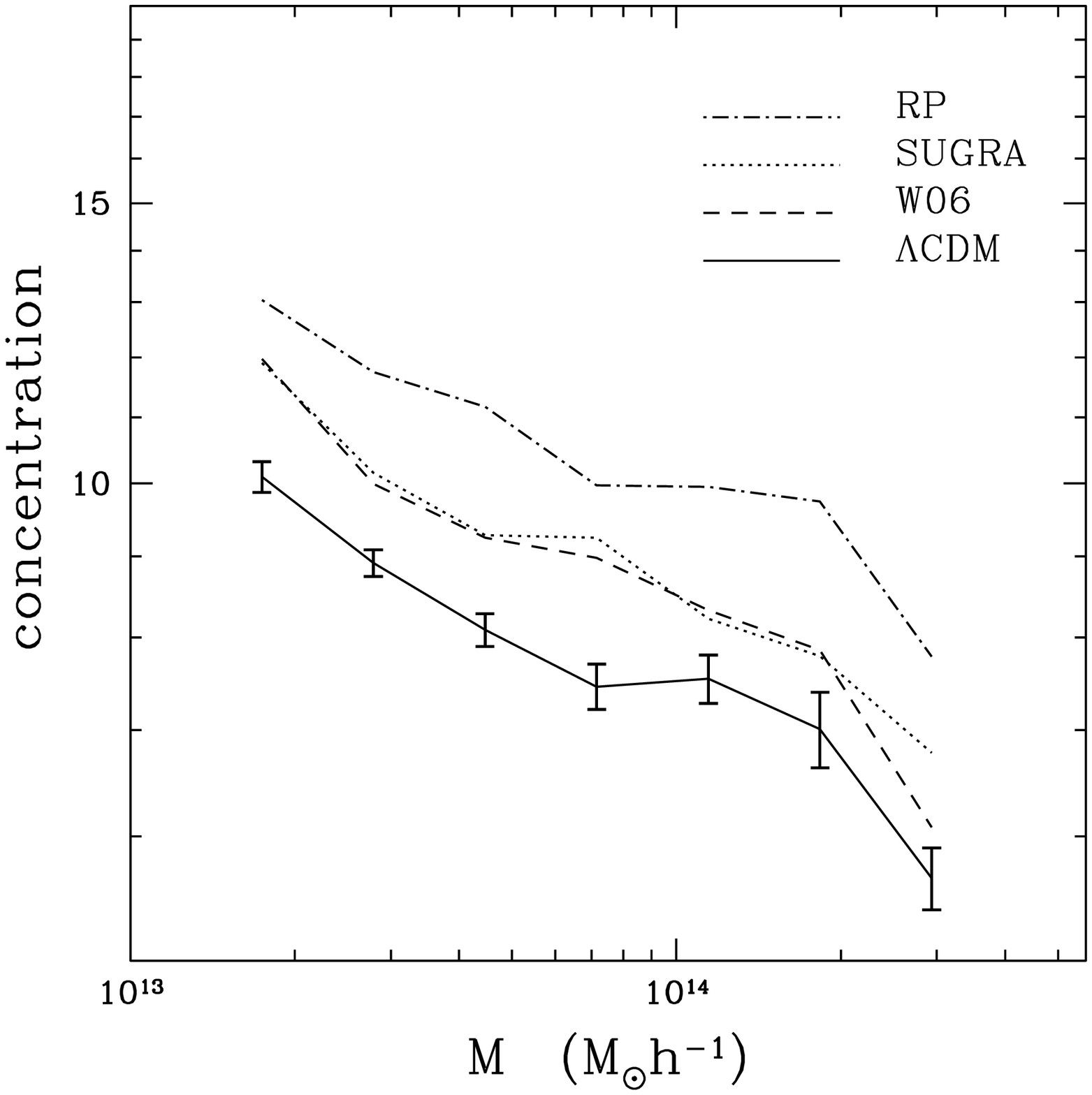}}
\caption{Concentration distribution in various models }}
\end{figure}

Using ART facilities, a particular halo was magnified in all
simulations. Fig.~5 shows that its profile is NFW with a concentration
depending on DE nature. Concentration can also be considered
on a statistical basis. Fig. 6 shows how halo concentrations depend on
the model. Here concentrations are defined as the ratio between
the radius $r_c$ at which the density contrast is 110 and the radius
$r_s$ in the NFW expression of the radial density.

\begin{figure}[t]
{\centerline{\includegraphics[width=2in,height=1.8in]{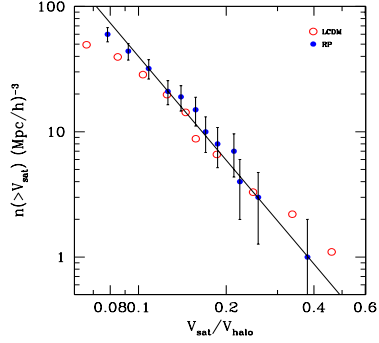}}
\caption{ Number of halo satellites }}
\end{figure}

We also studied how the number of satellites of a halo
depends on DE nature. In Fig.~7 we report such dependence.
However, also in this case, once
care is payed to properly normalize numbers to the same central halo
velocity, no appreciable dependence on DE nature can be found.

\section{Conclusions }
In this paper we showed how a simple modification of the program ART
 permits to perform a wide
analysis of dynamical DE models. This task is simplified by the very
structure of the program, which uses the scale factor $a$ as 
{\it time}--variable and requires only 
$$
dt/da = H_o^{-1} \sqrt{a \, \Omega_m(a)/\Omega_{m}(a_o)}
$$
($H_o$: today's Hubble parameter), to detail the action of forces. 
Once  $\Omega_m(a)$ (the dependence of the matter density 
parameter on the scale factor) is assigned, the dynamical problem is then 
properly defined.
Most of the preliminary work was then performed at the post--linear 
level. This provided us suitable expressions for the virial density contrast,
so that halos can be selected in the correct way in all models, and also the
required fitting expressions for $\Omega_m(a)$.

Discriminating DE models from $\Lambda$CDM essentially requires
good data at high redshift. A discrimination at $z=0$ can be
made only using an observable sensitive to the concentration distribution.
In principle, such an observable exists and is related to strong
lensing (giant halo statistic). Further work in this direction is in progress.


\begin{chapthebibliography}{1}



\bibitem{Bart98}
  Bartelmann M., Huss A., Carlberg J., Jenkins A. \& Pearce F.\
  1998, A\&A 330, 1
\bibitem{Bart02}
  Bartelmann M., Perrotta F. \& Baccigalupi C.
  2002, A\&A 396, 21
\bibitem{BraxMartin99}
  Brax, P. \& Martin, J., 1999, Phys.Lett., B468, 40
and 2000, Phys.Rev. D, 61, 103502
\bibitem{Efstathiou2}
  Efstathiou, G. et al., 2002, MNRAS, 330, 29
\bibitem{kl0}
Klypin, A., Maccio' A.V., Mainini R. \&  Bonometto S.A., 
2003, ApJ in press,  astro--ph/0303304 
\bibitem{kogut2003}
  Kogut et al., 2003, astro--ph/0302213
\bibitem{Mainini03a}
  Mainini R., Maccio' A.V. \& Bonometto S.A., 
  2003a, NewA 8, 172
\bibitem{Mainini03b}
  Mainini R., Maccio' A.V., Bonometto S.A., \& Klypin, A., 2003b,
  ApJ in press, astro--ph/0303303 
\bibitem{Netterfield} 
  Netterfield, C.~B.~et al.\ 2002, ApJ, 571, 604 
\bibitem{Percival}
  Percival W.J. et al., 2002, astro-ph/0206256, MNRAS (in press)
\bibitem{Perlmutter}
  Perlmutter S. et al., 1999, ApJ, 517, 565
\bibitem{Pogosian03}
  Pogosian, D., Bond, J.R., \& Contaldi, C. 2003, astro-ph/0301310
\bibitem{RP} 
  Ratra B., Peebles P.J.E., 1988, Phys.Rev.D, 37, 3406
\bibitem{Riess}
  Riess, A.G. et al., 1998, AJ, 116, 1009
\bibitem{spergel2003}
  Spergel et al.\  2003, astro--ph/0302209
\bibitem{Tegmark01} 
  Tegmark, M., Zaldarriaga, M., \& Hamilton, AJ\ 2001, Phys.R., D63, 43007 
\bibitem{wett1}
  Wetterich C., 1985, Nucl.Phys.B, 302, 668

\end{chapthebibliography}
\end{document}